\documentclass{ieeeaccess}

\usepackage{lineno,hyperref}
\usepackage{bm}
\usepackage{bbold}

\usepackage[dvipsnames]{xcolor}
\usepackage{xspace,comment}

\usepackage{cite}
\usepackage{amsmath,amssymb,amsfonts}
\usepackage{algorithmic}
\usepackage{graphicx}
\usepackage{textcomp}
\def\BibTeX{{\rm B\kern-.05em{\sc i\kern-.025em b}\kern-.08em
    T\kern-.1667em\lower.7ex\hbox{E}\kern-.125emX}}
\begin{document}
\history{Date of publication xxxx 00, 0000, date of current version xxxx 00, 0000.}
\doi{10.1109/ACCESS.2017.DOI}

\title{Towards Model Reduction for Power System Transients with Physics-\\Informed PDE}
\author{\uppercase{Laurent Pagnier}\authorrefmark{1}, \IEEEmembership{Member, IEEE},
\uppercase{Julian Fritzsch}\authorrefmark{2,3}, \IEEEmembership{Graduate Student Member, IEEE}, \uppercase{Philippe Jacquod}\authorrefmark{2,3}, \IEEEmembership{Member, IEEE}, and Michael Chertkov\authorrefmark{1},
\IEEEmembership{Senior Member, IEEE}.}
\address[1]{Program in Applied Mathematics, University of Arizona, Tucson, United States.}
\address[2]{School of Applied Sciences of Western Switzerland HES-SO, Sion, Switzerland.}
\address[3]{Department of Quantum Matter Physics, University of Geneva, Geneva, Switzerland.}
\tfootnote{This work has been partially supported by the Swiss National Science Foundation under grant 200020\_182050.}

\markboth
{Author \headeretal: Preparation of Papers for IEEE TRANSACTIONS and JOURNALS}
{Author \headeretal: Preparation of Papers for IEEE TRANSACTIONS and JOURNALS}

\corresp{Corresponding author: Laurent Pagnier (e-mail: laurentpagnier@math.arizona.edu).}

\begin{abstract}
This manuscript reports the first step towards building a robust and efficient model reduction methodology to capture transient dynamics in a transmission level electric power system. Such dynamics is normally modeled on seconds-to-tens-of-seconds time scales by the so-called swing equations, which are ordinary differential equations defined on a spatially discrete model of the power grid. Following Seymlyen (1974) and Thorpe, Seyler, and Phadke (1999), we suggest to map the swing equations onto a linear, inhomogeneous  Partial Differential Equation (PDE) of parabolic type in two space and one time dimensions with time-independent coefficients and properly defined boundary conditions. We illustrate our method on the synchronous transmission grid of continental Europe. We show that, when properly coarse-grained, i.e., with the PDE coefficients and source terms extracted from a spatial convolution procedure of the respective discrete coefficients in the swing equations, the resulting PDE reproduces faithfully and efficiently the original swing dynamics. We finally discuss future extensions of this work, where the presented PDE-based modeling will initialize a physics-informed machine learning approach for real-time modeling, $n-1$ feasibility assessment and transient stability analysis of power systems.  
\end{abstract}


\titlepgskip=-15pt

\maketitle

\section{Introduction}\label{sec:intro}

This manuscript is focused on building a computationally efficient and sufficiently accurate model describing the transient response of a transmission level electric power system to a significant perturbation -- for example the disconnection and/or reconnection of a large generator. We consider the dynamics of the transmission level of power systems on a continental scale and focus on sub-minute transients on time scales ranging from one second to few tens of seconds. We follow an approach that is standard in power system studies and assume that the so-called swing equations \cite{kundur,machowski1997power}, giving the time-evolution of the voltage angles at all nodes on the power grid, provide a sufficiently accurate representation of the power system dynamics within the considered spatio-temporal scales. Stated differently in the language of modern machine learning, the spatio-temporal integration of the swing equations provide a high-fidelity representation of the ground truth. There are two competing aspects of the swing equations.  On the one hand, they are based on physically meaningful quantities and parameters such as line capacities, machine inertia and damping. Accordingly they are expected to correctly capture the physics of the system. On the other hand, integrating these equations on a large, continental scale grid can be computationally very expensive, even for a single run corresponding to a specific initial condition. Obviously, it becomes even more expensive if the task is to screen many possible initial conditions, and often prohibitively expensive when the screening need to be repeated numerous times, testing many possible control actions. 
Model reduction for this type of {\it online} applications \cite{online-convex-optimization} comes as a way to strike a balance between accuracy and computational complexity. Central to this optimization is that the transient dynamics of interest occurs over time scales up to few tens of seconds while the goal is to
numerically resolve multiple scenarios of initial conditions and various controls faster than real time. 

{\bf How does model reduction work?} In the current era of deep learning, many model reduction techniques rely on neural networks and other tools of modern data science and machine learning, see e.g. \cite{ma2018model,2019Swischuk,bhattacharya2021model,2021Chen}. The idea is to use the ground truth model -- the swing equations in our case -- to produce dynamical data, and then to train a pre-selected reduced model on these datasets to fine-tune the parameters of the model. If the reduced model is of an application agnostic type, as is customary in mainstream machine learning, the scheme relies on very large datasets. However, recall that running our ground truth model is computationally expensive. Then, if producing the needed training datasets is not an option, 
can we still hope to build a reliable reduced model? Our only hope in this case is to inject the relevant, application-specific information -- in our case information about the power system physics -- into the model reduction framework. Physics-Informed Machine Learning (PIML) is the modern approach to resolve the model reduction bottleneck -- that is to compensate for the lack of data (typical of online applications) by building models that are aware of the underlying physics \cite{king2018deep}, as, e.g., expressed in terms of differential equations  \cite{Lagaris_nn_ode98,raissi2,21PIML-George}. (See also \cite{misyris2020physicsinformed,pagnier2021physicsinformed} for discussion of the application of PIML to power systems.) 

{\bf Why is Partial Differential Equation (PDE) modeling a sound option for power system model reduction?}
In this manuscript, we propose a first step towards developing PIML
for general online applications and
 advancing model reduction of PIML applications to power systems, as e.g. developed earlier by members of our team \cite{pagnier2021physicsinformed,pagnier2021embedding}. 
Similarly to \cite{misyris2020physicsinformed}, we take advantage of the PIML approach and construct an online framework for simulating power system dynamics faster than real time. We are however aiming to capture the transient dynamics in a very large, continental-scale power system,
a goal that has not been addressed by any earlier 
related approach we are aware of. Accordingly, we choose to build our reduced model on the continuous PDE approach to modeling power system dynamics pioneered by Semlyen \cite{semlyen1974analysis} and later extended by Thorpe, Seyler, and Phadke \cite{thorp1998electromechanical}   (see also \cite{parashar2004continuum}). These works were however restricted to spatially-continuous one dimensional, time-dependent systems, i.e. with 1+1 dimensional PDE. Our PDE approach to power systems, to be presented below, inherits all the relevant physics of the original swing Ordinary Differential Equations (ODEs), accordingly it is 2+1 dimensional. Thus, it resolves power grid dynamics over a spatially-continuous two-dimensional domain associated with the power system's geographical area of service. Approximating the swing ODEs by a PDE may seem strange at first sight, as naively, this transition dramatically increases the number of degrees of freedom. However, this naive thinking is not quite right for several reasons. First, numerical solutions of linear 2+1 dimensional PDE assume spatial regularization via a two-dimensional grid, where the grid size can be chosen according to the desired spatial resolution. Therefore, the number of grid points may eventually be comparable to or even smaller than the number of nodes in the original grid. Second, numerical operations, such as matrix inversion, can be performed much more efficiently on a regular grid than on a complex
meshed graph. Third, and most importantly, the number of physical parameters in the original power grid model (line capacities, machine inertia and damping coefficients) may be reduced significantly within the PDE approach. Indeed, within the reduced model, we want to faithfully capture only the long-wavelength components of the swing dynamics. This justifies using a coarse-grained/filtered expression for all the coefficients in the linear PDE, therefore representing the coefficients via only a few long-wavelength harmonics.

{\bf Our Contribution:} In this manuscript we make the first steps towards a novel online methodology for multi-scenario testing and control based on modeling the dynamics of a large, continental scale power system within a novel 2+1 PDE modeling framework. We show how a
properly coarse-grained PDE model 
faithfully captures the power grid transient responses to disturbances of a high fidelity model.  
Our approach is rather heuristic, but it is backed by physical understanding of how perturbations propagate over the grid. Therefore, we ``prove by example'' -- illustrating our model reduction methodology
on the PanTaGruEl model
of the synchronous grid of continental Europe introduced in \cite{pagnier2019inertia,tyloo2019key}. Specifically, PanTaGruEl 
simulates power flow and swing equations with high fidelity
to produce the ground truth data. The latter in their turn are used to infer a spatially continuous 2+1 dimensional PDE model. The quality of the reconstruction is judged, first, by its ability to mimic power system dynamics 
 and, second, by a faithful 
 reconstruction of spatially coarse-grained and physically meaningful static -- spatial distribution of line impedances -- and dynamic -- spatial distribution of damping and inertia -- parameters.
We conclude the manuscript with a suggestion for a path towards using the PDE based reduced modeling framework for efficient online screening of multiple failure scenarios on large transmission grids, faster than real time. 

\section{Problem Formulation}

\subsection{Power Flow and Swing Equations (system of ODEs)}

AC Power Flow (PF) equations describe steady distributions of electric power flows over an AC power grid. The equations connect complex power injections $\{ s_i\equiv p_i+\mathfrak{i}\, q_i\}$ to complex voltages $\{{V}_i\equiv v_{i}\exp(\mathfrak{i}\, \theta_i)\}$, where $p_i$, $q_i$, $v_i$ and  $\theta_i$ denote the active and reactive power injections, and the voltage magnitude and angle at node $i\in\mathcal{V}$ respectively:
\begin{subequations}\label{eq:load_flow2}
\begin{align}\label{eq:load_flow2a}
p_i&=\sum_{j}v_iv_j\Big[g_{ij}\cos\big(\theta_i-\theta_j\big)+b_{ij}\sin\big(\theta_i-\theta_j\big)\Big],\!\!\!\!\\\
\label{eq:load_flow2r}
q_i&=\sum_{j}v_iv_j\Big[g_{ij}\sin\big(\theta_i-\theta_j\big)-b_{ij}\cos\big(\theta_i-\theta_j\big)\Big].\!\!\!\!
\end{align}
\end{subequations}
Here, $b_{ij}$ and $g_{ij}$ are elements of the susceptance and conductance matrices, see e.g. \cite{bergen2000power} for more details.

Suppose that a steady solution of the PF Eqs.~(\ref{eq:load_flow2}) is perturbed, for example by a fast disconnection and reconnection of a large generator or load.  Such a fault induces a transient voltage angle and amplitude dynamics. It is customary to assume a time-scale separation between voltage amplitudes and angles. On time scales ranging from sub-seconds to few tens of seconds, voltage amplitudes remain constant, while voltage angles evolve according to the swing equations~\cite{machowski1997power,vancutsem1998voltage},
\begin{equation}\label{eq:swing}
m_i\ddot\theta_i+d_i\dot \theta_i=p_i-\sum_j v_i v_j b_{ij}(\theta_i-\theta_j)\,.
\end{equation}
The voltage amplitudes $v_i$ and $v_j$ are considered constant, already stabilized to the
steady-state solution of Eqs.~\eqref{eq:load_flow2} and $m_i$ and $d_i$ denote the inertia and the damping (i.e., primary control) of the generators.
\begin{figure}
\center
\includegraphics[width=\columnwidth]{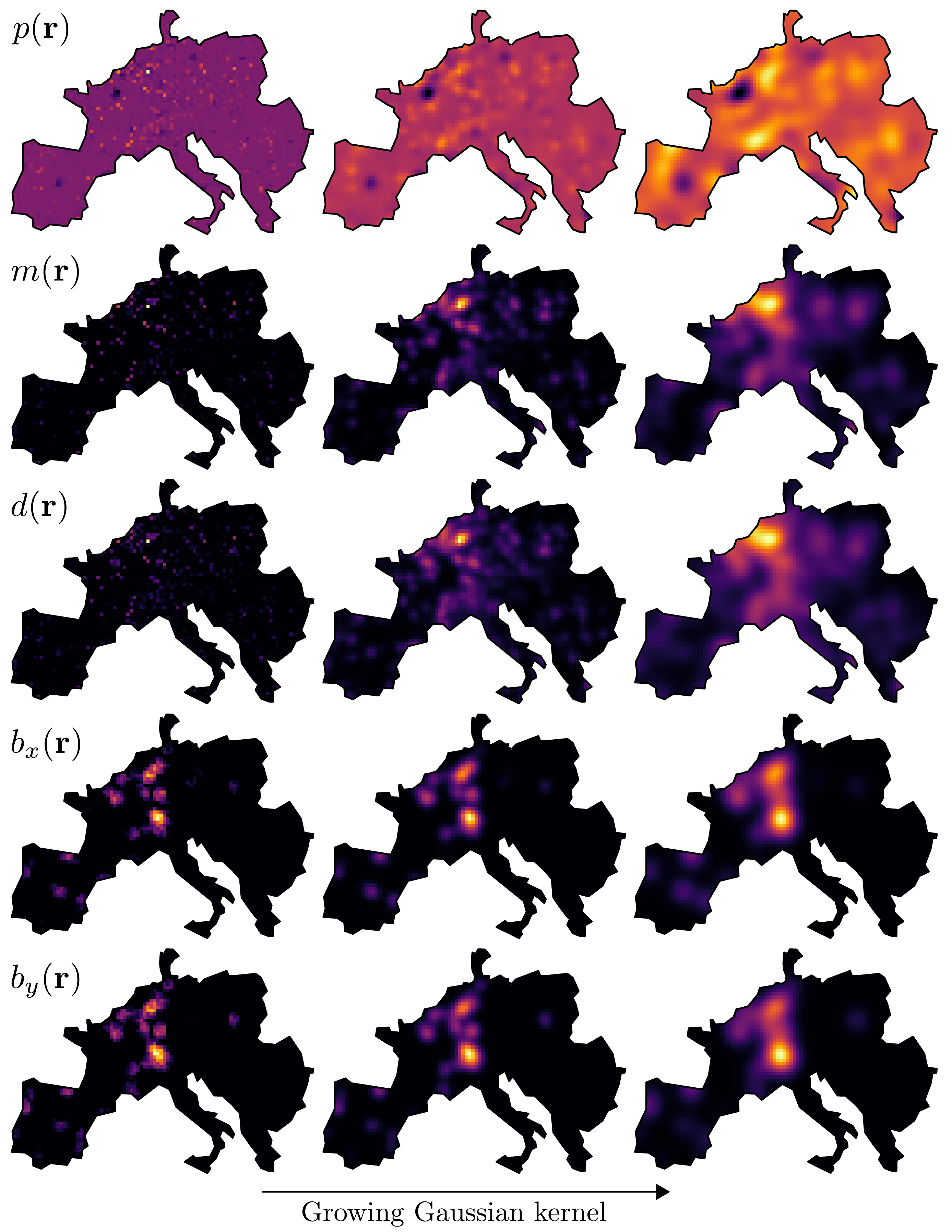}
\caption{Evolution of the system parameters for the continuous PDE model with increasing Gaussian kernel. (See text for details.) 
}
\label{fig:diff}
\end{figure}
Eq.~(\ref{eq:swing}) describes the relaxation dynamics of voltage angles towards a steady-state solution, $\ddot\theta_i=\dot\theta_i=0$, 
corresponding to the lossless, $g_{ij}=0$,
linearized version of Eqs.~(\ref{eq:load_flow2a}). Two comments are in order here. First, the linearized approach used here is in practice quite accurate to reproduce the transient dynamics following not too strong perturbations - a problem called small signal stability~\cite{machowski1997power}. Nevertheless, the approach to be presented below can be extended to the nonlinear case, with $(\theta_i-\theta_j) \rightarrow \sin(\theta_i-\theta_j)$ in Eq.~\eqref{eq:swing}. Second, 
the swing equation approach is not restricted to the just discussed case of a fast disconnection-reconnection fault, but also captures the voltage angle dynamics following a fault which is not immediately cleared, such as the 
removal of a generator or a load without reconnection. In such cases, the final relaxed state is not balanced, i.e. $\sum_i p_i\neq 0$, and the power mismatch is compensated by the second, damping term in Eq.~\eqref{eq:swing}, leading to an AC frequency shift   $\dot\theta_i=\omega_{pf} \, \forall i $,
with $\sum_i p_i=\omega_{pf} \sum_i d_i$, where $\omega_{pf}$ denotes the post-fault synchronous frequency.

In the next paragraph, we construct a reduced model by
mapping the discrete system of ODEs \eqref{eq:swing} into a
continuous PDE. Before we do that, we 
re-emphasize why a model reduction is needed at all. 
The motivation was lucidly expressed in Ref.~\cite{99Parrilo} as follows: ``The focus is on the construction of low-order models which closely approximate the global behavior of the hybrid nonlinear system. There is a growing recognition of the strong need
for rapid and reliable computation of the system dynamics." 
Comprehensive discussions of model reduction in a general context as well as for specific applications to slow coherency and inter-area oscillations can further be found in Ref.~\cite{chow2013power}.

\subsection{Mapping the power system to a two-dimensional continuum: the Swing PDE Model}
\label{subsec:PDE}

Consider a two-dimensional domain ${\bm \Omega}\subset \mathbb{R}^2$,
with coordinates ${\bm r}=(x,y)$, inside which the discrete, planar or quasi-planar network
is embedded. The boundary of the domain is denoted by $\bm{\partial \Omega}$ and $\forall {\bm r}\in \bm{\partial \Omega}$, $\bm n\equiv (n_x, n_y)$ denotes the normal vector to the boundary at ${\bm r}$.
Imagine that the swing Eqs.~(\ref{eq:swing}) are derived by discretizing a PDE describing the dynamics of a scalar field $\theta(t;{\bm r})$ over an irregular mesh which corresponds to the original network. Then, following \cite{thorp1998electromechanical}, one naturally asks:  what is the PDE corresponding to the swing Eqs.~(\ref{eq:swing})? We answer this question by writing the following, most general form of the swing PDE on ${\bm \Omega}$:
\begin{align}
   m(\bm r)\frac{\partial^2}{\partial t^2}\theta(t;\bm r)&+d(\bm r)\frac{\partial}{\partial t}\theta(t;\bm r)=p(t;\bm r)\nonumber\\
&+ \sum\limits_{\alpha,\beta=1,2}\partial_{{ r}_\alpha} b_{\alpha\beta}({\bm r})\partial_{{r}_\beta} \theta(t;\bm r),\label{eq:pde}
\end{align}
where ${r}_1=x,\ {r}_2=y$. One of our main task is to map 
the physical parameters of \eqref{eq:swing} into 
the continuum as follows
\begin{align}
\forall i:\quad  &\theta_i(t) \rightarrow \theta(t;\bm r),\ m_i \rightarrow m(\bm r),\  d_i \rightarrow d(\bm r)\nonumber\\
&p_i(t) \rightarrow p(t;\bm r),\   b_{ij} \rightarrow b_{\alpha\beta}(\bm r),\ \forall \alpha,\beta =1,2.
\end{align}
We discuss a procedure for initializing these continuous parameters in Section~\ref{sec:AD}.

\begin{figure}
\center
\includegraphics[width=\columnwidth]{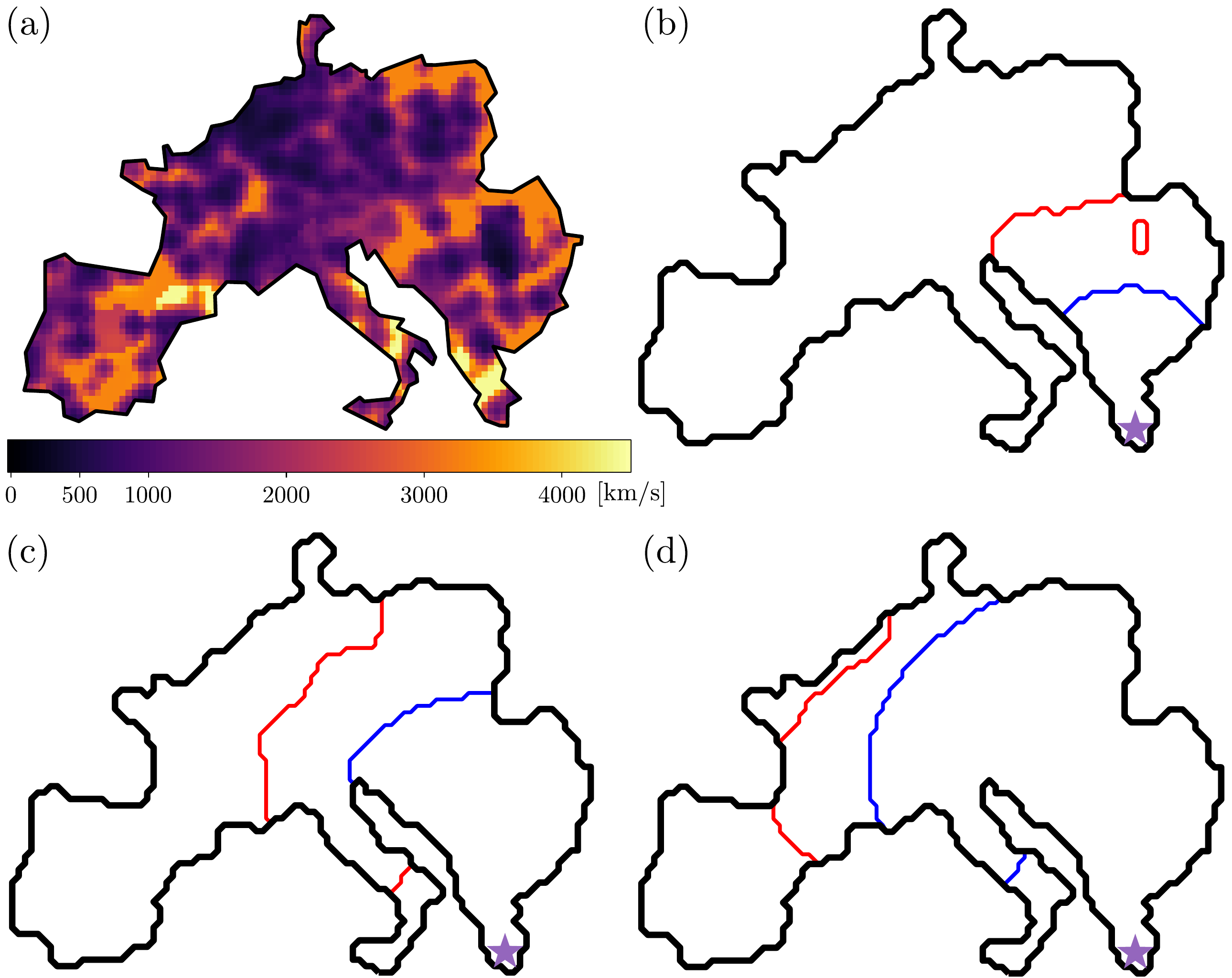}
\caption{(a) Assessment of the local propagation speed as $c({\bm r})=\sqrt{b({\bm r})/m({\bm r})}$. (b)-(d) Fronts of the perturbation at incremental time intervals of $\Delta t=0.6$s, after a fault in Greece (violet star), for inhomogeneous (red) and average parameters (blue).}
\label{fig:prop}
\end{figure}

Next, the swing PDE (\ref{eq:pde}) must be constrained with physically
appropriate boundary conditions. In our case, they are Neumann 
boundary conditions
\begin{align}\label{eq:Neumann}
    \forall t,\ \forall {\bm r}\in {\bm \partial\Omega}:\ \sum\limits_{\alpha,\beta=1,2}n_\alpha(\bm r) b_{\alpha\beta}(\bm r) \partial_{{r}_\beta}\theta(t;{\bm r})=0,
\end{align}
corresponding to a vanishing normal derivative of the angle field on
the domain boundary $\bm{\partial\Omega}$. These boundary conditions 
directly follow from the condition that 
post-perturbation frequencies in the continuous model correspond to those in the original swing equations, i.e. $\omega(t;\bm r)\equiv\frac{\partial}{\partial t}\theta(t;\bm r) = \omega_{\rm pf}\,,\;\forall \bm r\in \Omega$. This condition translates into
\begin{equation}
\begin{split}\label{eq:bc}
\omega_{\rm pf} \intop_{\bm \Omega}d(\bm r){{\rm d}\bm r} &= \intop_{\bm \Omega} p(t; \bm r){{\rm d} \bm r}\\
&+\sum\limits_{\alpha,\beta=1,2}\;\intop_{\bm {\partial\Omega}} n_\alpha({\bm r})b_{\alpha\beta}(\bm r) \partial_{{r}_\beta}\theta(t;\bm r)  {{\rm d}\bm r}\, .
\end{split}
\end{equation}
This directly corresponds to the frequency shift
$\sum d_i\omega_{\rm pf}=\sum_i p_i$ in the swing Eqs.~\eqref{eq:swing}
if the second term in the right-hand side of Eq.~(\ref{eq:bc}) 
vanishes identically, which is guaranteed by the Neumann boundary conditions (\ref{eq:Neumann}).

In the following we simplify our PDE model, assuming that the $b$-tensor is diagonal $b_{12}=b_{21}=0$, accordingly, we use a shorter notation, $b_{11}\to b_x,\ b_{22}\to b_y$. We will shortly show that it can be even more simplified to $ b_y(\bm r)\approx b_x(\bm r) =:b(\bm r)$.

\begin{figure}
\includegraphics[width=\columnwidth]{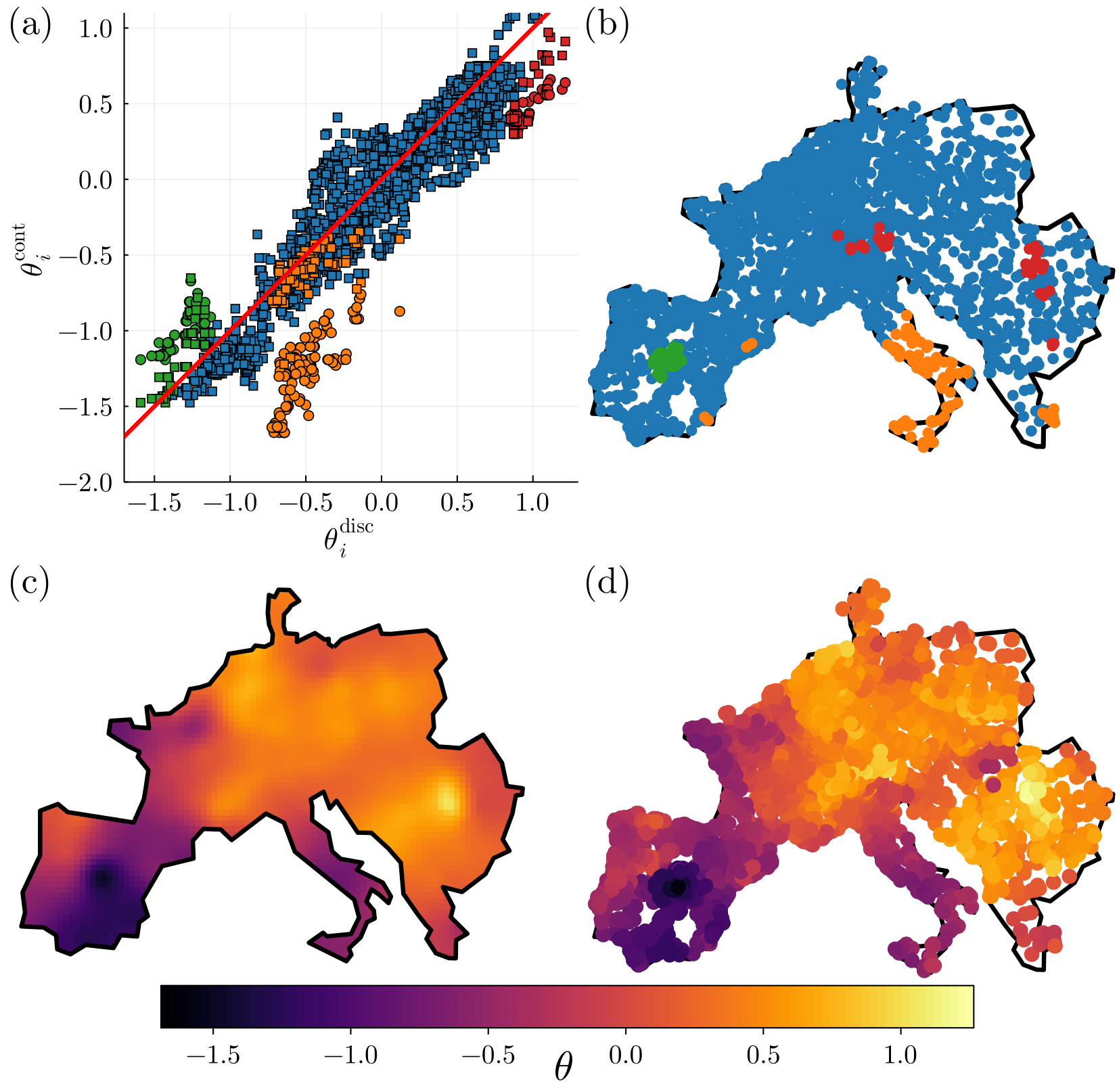}
\caption{Comparison of ground-truth and continuous PDE steady state solutions. (a) One-to-one comparison of local voltage angle: for each bus in the discrete model the  nearest node in the continuous mesh is selected. The red line indicates a perfect match. The outliers marked in orange, red and green correspond to the points marked on the map in (b). The square markers correspond the solution after adjusting the susceptances as outlined in the text. (c) Continuous solution $\theta(\bm r)$ after adjustment. (d) Discrete solution $\theta^{\rm disc}$.}
\label{fig:steady}
\end{figure}

\section{Initialization of PDE Parameters and calibration}\label{sec:AD}

Once the general structure of the swing PDE (\ref{eq:pde}) is established, we need to initialize the PDE parameters $b_{x,y}({\bm r}), m({\bm r})$ and $d({\bm r})$.  This is achieved by applying a smoothing to the original parameters defined on the discrete grid of Eq.~\eqref{eq:swing}.
Obviously, there are many different choices for this coarse-graining/filtering procedure. Therefore, it is crucial to develop a validation criteria.
We calibrate and validate via post factum tests, described in the following section, where we compare the 
dynamics of a fault in the original, discrete swing equations with that in the spatially continuous model. In future work, this initialization procedure will be complemented by a machine learning scheme performing tailored adjustments to increase the accuracy of the model even more.

This simple smoothing procedure was proposed in Ref.~\cite{semlyen1974analysis, thorp1998electromechanical} focusing on 
a 1+1 dimensional PDE representation of a linear power network, where all parameters in the 1+1 dimensional PDE were chosen to 
be spatially constant. This homogeneous smoothing procedure was improved in \cite{parashar2004continuum}, where non-uniform parameters of the 1+1 PDE system were derived by means of a convolution with a fixed Gaussian kernel.  

We introduce a slight generalization of this Gaussian smoothing process. We apply an Artificial Diffusion (AD) to spatial distributions of the physical coefficients, $m({\bm r}), d({\bm r})$ or $b_{\alpha\beta}({\bm r}$). It starts by assigning discrete physical quantities to the nearest nodes discretizing the PDE (\ref{eq:pde}), then they diffuse over the lattice. The longer the diffusion is allowed, the broader the convolution kernel. The diffusion is stopped when parameters satisfy some smoothness criterion. This generalization is advantageous because it allows the optimal width of the Gaussian kernel to be self-determined (no additional criteria are required).

 \begin{figure*}[t]
\includegraphics[width=\textwidth]{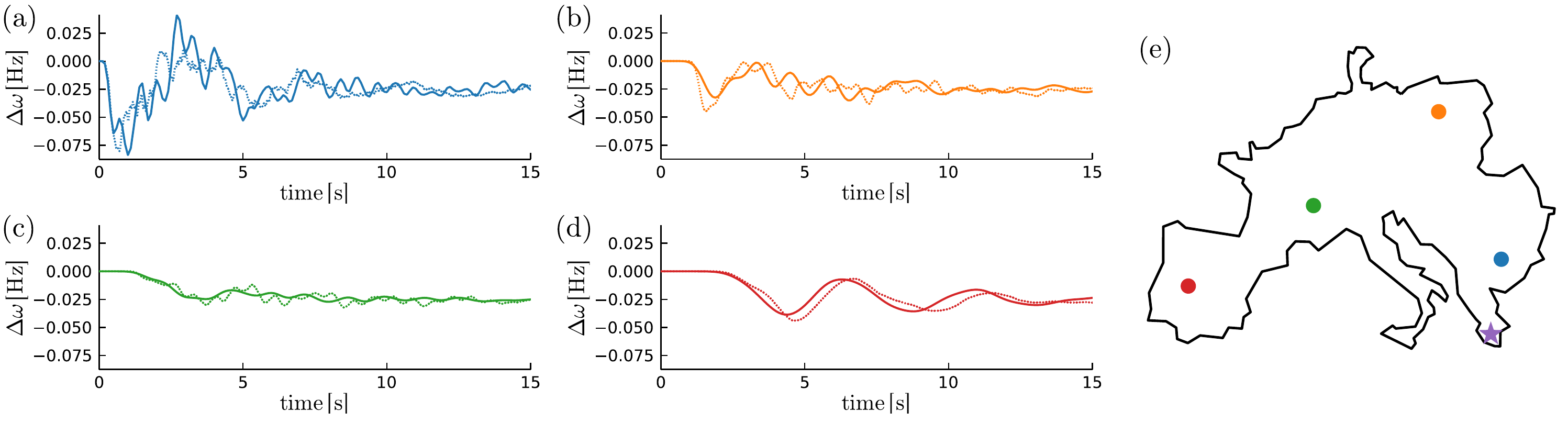}
\caption{Frequency response of generators in (a) Bulgaria, (b) Poland, (c) France, and (d) Spain to a 900~MW loss of power in Greece. Data from the continuous PDE and the ground-truth ODE models are displayed as dotted and solid lines respectively. The map (e) shows the locations of the fault (marked by the purple star) and of the generators considered in panels (a)-(d) (marked by dots with the color corresponding to the respective frequency plots). }
\label{fig:dyn}
\end{figure*} 

We illustrate the AD process on PanTaGruEl which consists of $3809$ buses, $618$ generators and $4944$ lines. There are $3221$ nodes in the discretization of our continuous model. Even though this number is not significantly smaller than the number of buses in the discrete system, the continuum model can  be parametrized efficiently with a much fewer parameters than what is required in the discrete model as we will show shortly.
Fig.~\ref{fig:diff} illustrates the AD process on the susceptances, the damping and inertia coefficients and the power injections. 
These numerical results suggest in particular  that the resulting spatial distribution of the diagonal part of the susceptance tensor is isotropic, i.e. $b_x(\bm r)\approx b_y(\bm r)$.  This, combined with the assumption that the off-diagonal terms of the $b$-matrix are much smaller than the diagonal ones (thus dropped), means that the entire susceptance tensor may be approximated by a scalar function, $b_{\alpha\beta}(\bm r)\approx \delta_{\alpha\beta} b({\bm r})$, where  $\delta_{\alpha\beta}$ is the Kronecker symbol.

We conclude this section by showing how already this relatively simple initialization procedure provides a useful and intuitive insight into the system behavior. Indeed, once the continuous values of the susceptance tensor, ${\bm b}({\bf r})$ and  of the inertia vector, ${\bm m}({\bm r})$, are determined and validated, we can immediately use them to build a spatial map of the electro-mechanical wave velocity, $c({\bm r})=\sqrt{b({\bm r})/m({\bm r})}$,  shown in Fig.~(\ref{fig:prop}). This figure also illustrates  how knowing the velocity, $c({\bm r})$, allows us to reconstruct the dynamical spread of a localized perturbation. Notice, however that these illustrations, even though encouraging, also suggest that one needs to be careful in extending the approach to the grids with strong degree of heterogeneity.  The extension is certainly possible,  however to achieve an accurate representation of the actual grid dynamics by the PDE model will require introducing more trainable parameters representing higher degree of spatial inhomogeneity.

\section{Numerical Validation} \label{sec:experiments}

In this Section, we juxtapose our PDE model, with the parameters found through the AD process explained in the preceding section, against the original swing model considered as the \textit{ground truth}. Therefore, our first validation task is to compare the steady state solution (of the static PF equations) with the solution of a Poisson problem associated with the static version of the PDE model. We then compare responses, within our PDE model vs the ground truth model, to a sufficiently large perturbation - a power outage. Finally, we show that the number of parameters describing the PDE model can be reduced dramatically without loss in accuracy. 

\subsection{Steady State Experiments}\label{sec:steady}

We start with a comparative analysis of the steady-state solutions of the PDE and of the ground-truth model. To do that, we find the grid points closest to the location of the buses within PanTaGruEl and terminate the AD process when the voltage angles  of the two steady-state solutions are as close to one another as possible, $\theta_i^{\rm cont} \simeq \theta_i^{\rm disc}$.  Comparison of the two solutions is shown in Fig.~\ref{fig:steady} (a). They are clearly in a good agreement overall, even though not without some  discrepancies. The outliers were highlighted in different colors  in Fig ~\ref{fig:steady} (b). We conjecture that the discrepancies are largely due to misrepresentation of parameters in the part of the grid with strong aspect ratio, e.g. at the Italian peninsula which is long and narrow. Specifically, in this case the boundary conditions we set in the continuous model may be too restrictive, effectively forcing parameters in the part of the grid with large aspect ratios to become much smaller than what we observe in the respective part of the discrete model. 

To verify the hypothesis, we simply modify susceptances in the parts of the grid corresponding to the  outliers. Specifically,  we increase susceptance uniformly within the Italian peninsula (as the aspect ratio there is large) and reduce it over the Iberian peninsula and Transylvania.  As seen in Fig.~\ref{fig:steady}, this simple and admittedly ad-hoc adjustment was sufficient to improve the agreement (the outlier effect was reduced significantly). We anticipate that a more accurate and automatic tuning (via ML tools which are work in progress) will produce even better results.

We conclude by mentioning that some of the discrepancies just discussed can  be attributed to transformers present in the ground truth model, but absent in the PDE model.  These and other strongly localized effects cannot be, properly represented in the continuous model. However, these discrepancies are expected to weaken at larger (spatial) scales -- that is at in the coarse-grained picture which is in the focus of our model reduction analysis. Moreover, we do not expect the transformers to play a significant role in analysis of transients, we are now switching our attention to.
\begin{figure*}
\center
\includegraphics[width=0.95\textwidth]{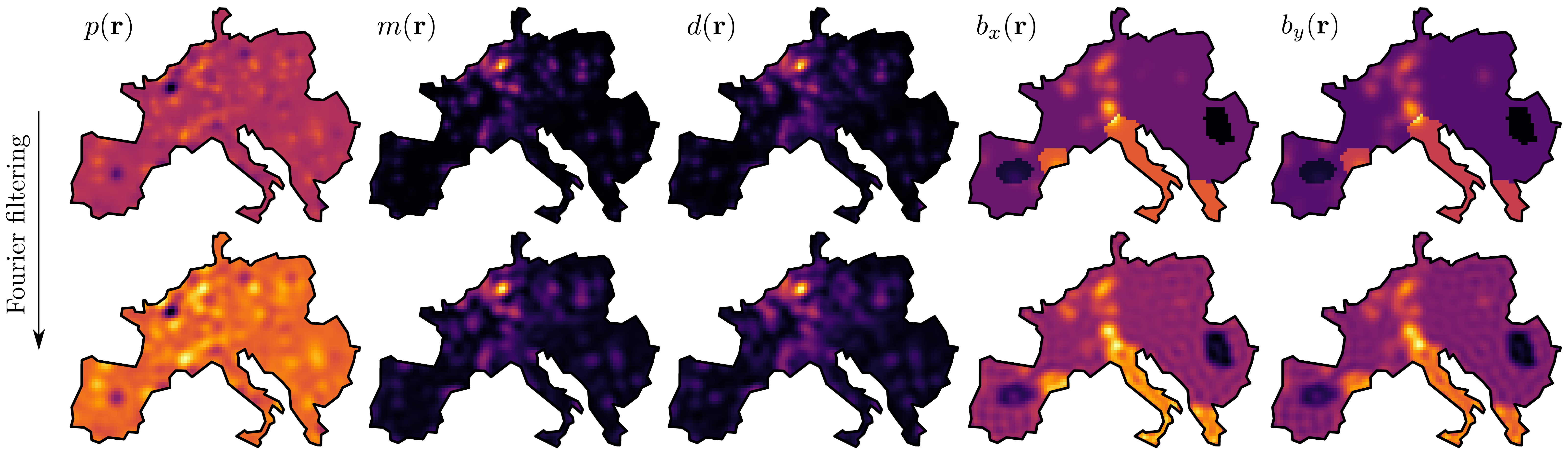}
\caption{Distribution of the grid parameters before and after a Fourier low pass filter with a cut-off frequency of 30\% (of the maximum frequency) applied.}
\label{fig:fourierparams}
\end{figure*}

\subsection{Dynamic Experiments}

Here we discuss how well the PDE model reproduces the ground truth dynamics observed in response to 
an abrupt removal of a 900~MW power plant in Greece. Fig~\ref{fig:dyn} shows comparison of the frequency response at four generators across Europe. The agreement between the reduced and ground truth model is very good. It is especially accurate far away from the fault. (We remind the reader that reproducing fatefully coarser picture, and not the small scale details, is exactly our goal.)
We also report that the propagation time and the frequency map of the inter-area oscillations are well reproduced too.
Not surprisingly, we also see that higher harmonics present close to the fault location (and gradually disappearing as we move away from the fault) are over-estimated by the PDE model. 
We expect that a more accurate -- machine-learning trained -- re-parametrization of the PDE model will be able to correct the problem, 
however on the expense of introducing higher-degree of parameter inhomogeneity in the PDE model.
In the end,  this is a matter of a trade-off decision which a designer of the reduced model should make --- this is a trade off between accuracy of prediction and degree of the model reduction.

\subsection{Moving towards Model Reduction}

As we mentioned earlier in the text, the number of harmonics in the parameters of the PDE model was set to be slightly less than the number  of  nodes in the original grid model.
Our next step is to 
see if we can reduce the number of harmonics in the PDE model coefficients  even further without any significant loss of accuracy in the spatio-temporal dynamics coarse-grained at the resolution of 50 km or larger. In other words we now study how our PDE model performs once we apply a low-pass filter to the coefficients.
\begin{figure}
\includegraphics[width=\columnwidth]{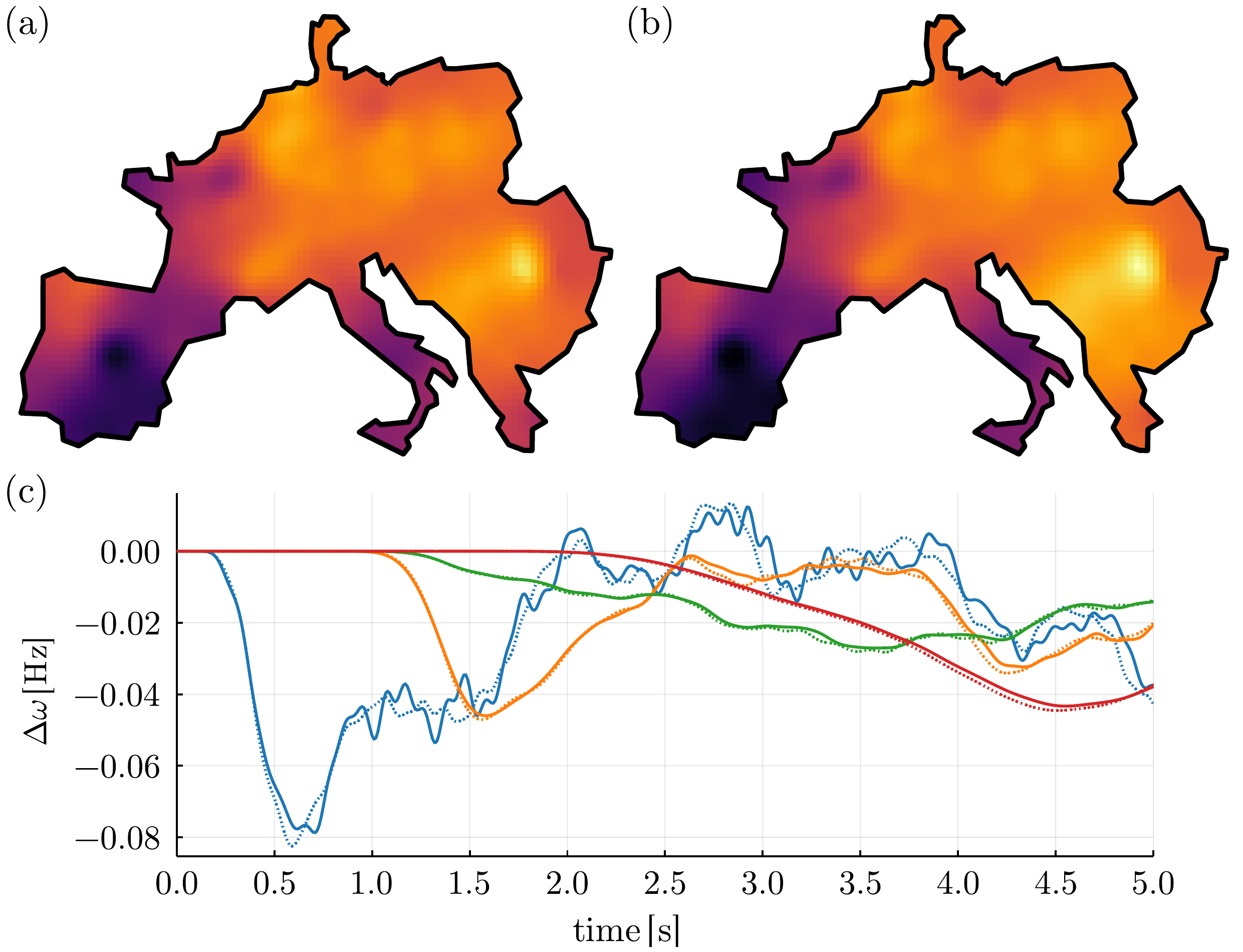}
\caption{Comparison of the steady state solution before (a) and after (b) application of the Fourier low-pass filter. The two solutions are in good general agreement, with some deviations in peripheral regions, e.g. in the Balkans or Spain. The same color scale as in Fig.~\ref{fig:steady} is used in panel (a) and (b). (c) Frequency response of the system to same fault as Fig.~\ref{fig:dyn} (same configuration), frequencies obtained with the filtered model are displayed with dashed lines. }
\label{fig:fourier}
\end{figure}

To investigate this matter, we perform an additional Fourier filtering to the results of the AD procedure. Specifically,  we choose the 
cut-off spatial frequency equal to 30\% of the largest spatial frequency set in the bare (unfiltered) version of the PDE model. The results of this filtering experiments are shown in Figs.~(\ref{fig:fourierparams},\ref{fig:fourier}). We observe that some compression artifacts, similar to ``ripples'', are present, but they seem to show little to no effect on the system dynamics, see in particular Fig.~\ref{fig:fourier} (c). Here we would like to emphasize (again) that a much more accurate filtering can be achieved with a ML approach, however it is encouraging to see that with a relatively simple tuning of parameters we were able to achieve reproduction of the principal features with such a good quality, even though out filtering (parameter fitting) procedure was certainly not optimal. Let us also notice that the coarse adjustments, mentioned in Section \ref{sec:steady}, are clearly visible.

\section{Critical Evaluations, Discussions and Future Work}\label{sec:conclusions}

Our main accomplishment in this manuscript is the construction and the validation of the 
continuous PDE model of the swing equations. The construction included the accurate resolution of the boundary conditions and the development of an efficient and flexible parameter filtering procedure based Artificial Diffusion (AD) and Fourier filtering. The validation proceeded via static
and dynamic comparisons of the continuous PDE model and the original discrete model. 

We also made a number of interesting observations which are clearly preliminary. Comparing the computation times for dynamical simulations of the PDE model on a regular grid of size comparable to the size of the original discrete graph we found that the PDE model is faster by at least a factor of ten. This is consistent with what we stated in the introduction. We also observed that in many instances a significant filtering of the dynamical parameters through a rather large coarse-graining scale (via wide Gaussian kernels and Fourier filtering) does not impact the accuracy of the dynamics of voltage frequency waves on sufficiently large scales (hundreds of kilometers) and sufficiently long times (seconds). 

Finally, this manuscript's most important message is that the model reduction presented here is only the starting point for an upcoming PIML methodology, where the functional maps for $m({\bm r}), d({\bm r})$ and $b_{\alpha\beta}({\bm r})$ will be modeled as Neural Networks. Specifically, future work will focus on using the approach developed here as a warm start for learning physical parameters of the PDE model (\ref{eq:pde}). Indeed, we envision modeling the functional maps for, $m({\bm r}), d({\bm r})$ and $b_{\alpha\beta}({\bm r})$ as Neural Networks. Hence, the AD procedure is still expected to be useful to initialize the future PIML schemes.

Planning for this work, we are aiming for keeping training process of the Machine Learning schemes under control in terms of computation time -- consistently with the goal of making them capable of achieving the goal of evaluating in parallel multiple perturbation scenarios in the time which is comparable or faster than dynamic simulations of the ground truth (swing) model. 

\section*{Acknowledgment}
The authors thank Robert Ferrando and Christopher Koh for participating in the discussions which led to the manuscript.

\appendix

\section{Details on the discretization of the PDE and its numerical integration}
We use the same spatial increment $\Delta$ for x and y axes, subsequently $\bm r=(i\Delta ,j\Delta)$.
\begin{align*}
b_x(\bm r)&=\frac{b^x_{i,j-1}+b^x_{i,j}}{2}\,,\\
\partial_xb_x&=\frac{b^x_{i,j}-b^x_{i,j-1}}{\Delta}+O(\Delta^3) \,,\\
\partial_x\theta&=\frac{\theta_{i+1,j}-\theta_{i-1,j}}{2\Delta}+O(\Delta^3)\,,\\
\partial_x^2\theta&=\frac{\theta_{i-1,j}-2\theta_{i,j}+\theta_{i+1,j}}{\Delta^2}+O(\Delta^4)\,.
\end{align*}
Similar expressions are obtained for the y-axis. Then discretization of the last term in Eq.\eqref{eq:pde} becomes
\begin{align}
\!\!\!&\partial_xb_x\partial_x\theta+b_x\partial_x^2\theta +\partial_yb_y\partial_y\theta+b_y\partial_y^2\theta\approx
\big(b^x_{i,j-1}\theta_{i,j-1}\nonumber\\
\!\!\!&+b^x_{i,j}\theta_{i,j+1}+b^y_{i-1,j}\theta_{i-1,j}+b^y_{i,j}\theta_{i+1,j}-\beta\theta_{i,j}\big)\!\Big/\!\Delta^2,\label{eq:operator}
\end{align}

where $\beta=b^x_{i,j-1}+b^x_{i,j}+ b^x_{i-1,j}+b^y_{i,j}$. In order to make our numerical scheme more efficient we vectorize (re-index) the grid, and the field, $\theta(t;{\bm r})$ defined over the grid, according to 
$\theta_{i,j} \rightarrow \tilde\theta_k$, where $k = N_y(i-1)+j$. It results in the following re-indexing of the grid-neighbors: 
$i-1,j \rightarrow k-1,\ 
i+1,j \rightarrow k+1, \
i,j-1 \rightarrow k-N_y, \
i,j+1 \rightarrow k+N_y$. 
This results in reformulation of the principal part of Eq.~\eqref{eq:operator} in terms of  a matrix $\bm \Xi$ acting on the vector $\bm{\tilde \theta}$. Furthermore, with the convention
that inner nodes, i.e. nodes that aren't on the boundary layer, have a zero normal vector, $n_x=0$ and $n_y=0$, and introducing $\eta_\pm(x) = \{1 \text{ if } \pm x\ge 0\,;\; 0 \text{ otherwise}\}$, we rewrite $\bm \Xi$, therefore accounting for the Neumann boundary conditions (\ref{eq:Neumann}),
\begin{align}
\Xi_{kl} = &-\tilde\beta_k\delta_{k,l}+\eta_+(n_x)\tilde b^{x}_{k-N_y}\delta_{k-N_y,l}
+\eta_-(n_x)\tilde b^{x}_{k}\delta_{k+N_y,l}\nonumber\\
&+\eta_+(n_y)\tilde b^{y}_{k-1}\delta_{k-1,l}
+\eta_+(n_y)\tilde b^{y}_{k}\delta_{k+1,l}\,,
\end{align}
where $\tilde\beta_k=\eta_-(n_x)\tilde b^x_{k}+\eta_+(n_x)\tilde b^x_{k-N_y}+\eta_-(n_y)\tilde b^y_{k}+\eta_+(n_y)\tilde b^y_{k-1}$ and $\delta_{\cdot,\cdot}$ is the Kronecker product. It is important that the method used for the numerical integration of the PDE is a finite volume method. This class of methods is conservative. This means that there is zero flux leakage at the boundary by construction which, in particular, guaranties that the post-fault system frequency is indeed at the value it is expected to be.

Finally, we use the Crank–Nicolson method \cite{crank1947practical} to integrate PDE~\eqref{eq:pde}. At  each time step we solve the following system of linear equations
\begin{equation}
\bm A\,
\left[\!\!\begin{array}{c}
\bm{\tilde \theta}(t+\Delta t)\\
\bm{\tilde \omega}(t+\Delta t)
\end{array}\!\!\right]= \bm B\,
\left[\!\!\begin{array}{c}
\bm{\tilde \theta}(t)\\
\bm{\tilde \omega}(t)
\end{array}\!\!\right]
 + \bm C\,,
\end{equation}
where
\begin{align*}
\bm A &= \left[\!\!\begin{array}{cc}
\mathbb{1} & -\frac{\Delta t}{2} \mathbb{1}\\
-\frac{\Delta t}{2}\,\bm M^{-1}\,\bm \Xi & \mathbb{1} + \frac{\Delta t}{2}\,\bm\Gamma
\end{array}\!\!\right]\,,\\
\bm B&=\left[\!\!\begin{array}{cc}
\mathbb{1} & \frac{\Delta t}{2} \mathbb{1}\\
\frac{\Delta t}{2}\,\bm M^{-1}\,\bm \Xi & \mathbb{1} - \frac{\Delta t}{2}\,\bm\Gamma
\end{array}\!\!\right]\,,\\
\bm C &= \left[\!\!\begin{array}{c}
\mathbb{0}\\
\frac{\Delta t}{2}\bm M^{-1}\Big(\bm{\tilde p}(t+\Delta t)+\bm{\tilde p}(t)\Big)\end{array}\!\!\right]\,,
\end{align*}
with $\bm M = {\rm diag}\big(\bm{\tilde m}\big)$ and $\bm \Gamma = {\rm diag}\big(\bm{\tilde m}^{-1} \bm{\tilde d}\,\big)$.

\bibliographystyle{IEEEtran}

\begin{IEEEbiography}[{\includegraphics[width=1in,height=1.25in,clip,keepaspectratio]{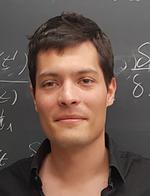}}]
    {Laurent Pagnier}(Member, IEEE) received the M.S. and Ph.D. degrees in theoretical physics from EPFL, Lausanne, Switzerland, in 2014 and 2019, respectively. He is a Visiting Assistant Professor at the University of Arizona. His research interest includes developing new modeling and monitoring methods for power systems.  He is currently aiming at applying Machine Learning techniques to power systems.  He is particularly interested in reinforcing the interpretablility and trustworthiness of ML methods which are paramount to increase their acceptance and usage inside this power system community.
\end{IEEEbiography}

\begin{IEEEbiography}[{\includegraphics[width=1in,height=1.25in,clip,keepaspectratio]{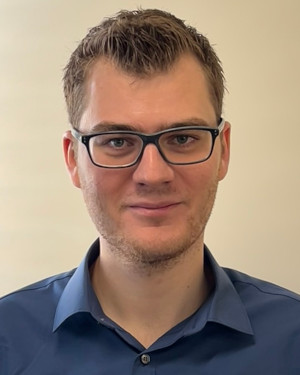}}]
    {Julian Fritzsch}(Graduate Student Member, IEEE) received the B.Sc. and M.Sc. degrees in physics from the University of Würzburg, Germany, in 2017 and 2020, respectively. He is currently pursuing the Ph.D. degree with the Department of Quantum Matter Physics, Univer- sity of Geneva, Switzerland, and the Electrical Energy Efficiency Group, University of Applied Sciences of Western Switzerland—HES-SO, Sion, Switzerland. His research interest includes dynamics of power systems.
\end{IEEEbiography}

\begin{IEEEbiography}[{\includegraphics[width=1in,height=1.25in,clip,keepaspectratio]{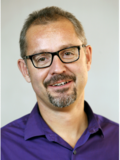}}]{Philippe Jacquod} (Member, IEEE) received the Diplom degree in theoretical physics from the ETHZ, Z\"{u}rich, Switzerland, in 1992, and the PhD degree in
    natural sciences from the University of Neuch\^{a}tel,
    Switzerland, in 1997. He is a professor with the Department of Quantum Matter Physics, University of Geneva, Switzerland and with the engineering department,
    University of Applied Sciences of Western Switzerland, Sion, Switzerland. From 2003 to 2005 he was an assistant professor with the
    theoretical physics department, University of Geneva,
    Switzerland and from 2005 to 2013 he was a professor
    with the physics department, University of Arizona, Tucson, USA. His main research topics is in power systems and how they
    evolve as the energy transition unfolds. He has published about 100
    papers in international journals, books and conference proceedings.
\end{IEEEbiography}

\begin{IEEEbiography}[{\includegraphics[width=1in,height=1.25in,clip,keepaspectratio]{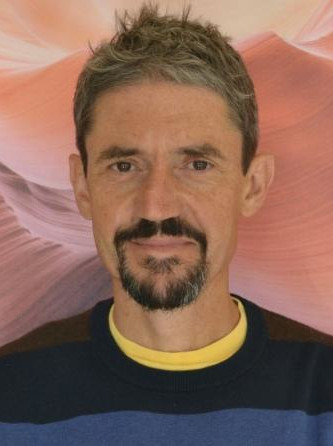}}]
    {Michael Chertkov }(Senior Member, IEEE) received the M.Sc. degree in physics from Novosibirsk State University, Novosi- birsk, Russia, in 1990, and the Ph.D. degree in physics from the Weizmann Institute of Science, Rehovot, Is- rael, in 1996. After his Ph.D., he spent three years at Princeton University as a R.H. Dicke Fellow in the Department of Physics. He joined Los Alamos Na- tional Lab in 1999, initially as a J. R. Oppenheimer Fellow in the Theoretical Division, and continued as a Technical Staff Member leading projects in physics of algorithms, energy grid systems, physics, and engi- neering informed data science and machine learning for turbulence. In 2019, he moved to Tucson to lead Interdisciplinary Graduate Program in Applied Math- ematics, University of Arizona, continuing to work for LANL part time. He has published more than 200 papers. His areas of interest include mathematics and statistics applied to physical, engineering, and data sciences. He is a Fellow of the American Physical Society.
\end{IEEEbiography}

\EOD

\end{document}